\documentclass[12pt,preprint]{aastex}

\usepackage{graphicx}
\usepackage{natbib}
\usepackage{amssymb}
\def\bsymbol#1{\mbox{\boldmath
$\displaystyle#1$\unboldmath}}

\newcommand*{\be}{\begin{equation}} 
\newcommand*{\ee}{\end{equation}}

\shorttitle{Neutrino mass spectrum from gravitational waves}

\shortauthors{Mosquera Cuesta \& Lambiase}

\begin{document}

\title{Neutrino mass spectrum from gravitational waves generated by double neutrino 
spin-flip in supernovae}
%%%  --- Date of preparation: 24 May 2008 --- 23:01' }

\author{Herman J. Mosquera Cuesta$^{1}$ and Gaetano Lambiase$^{2}$}

\affil{$^1$Instituto de Cosmologia, Relatividade e Astrof\'{\i}sica (ICRA-BR), 
Centro Brasileiro de Pesquisas F\'{\i}sicas (CBPF), Rua Dr. Xavier Sigaud 150, 
22290-180, RJ, Brazil}

% \affil{$^2$Abdus Salam International Centre for Theoretical
% Physics, Strada Costiera 11, Miramare 34014, Trieste, Italy}

\affil{$^2$Dipartimento di Fisica "E. R. Caianiello", Universit\'a
di Salerno, 84081 Baronissi (Sa), Italy. Also at INFN, Sezione di 
Napoli, Italy}

\date{Received 27 May 2008  --- 15:31'/ Accepted}

%---------------------------------------------------------------
\begin{abstract}
%---------------------------------------------------------------

The supernova (SN) neutronization phase produces mainly electron ($\nu_e$) 
neutrinos, the oscillations of which must take place within a few mean-free-paths 
of their resonance surface located nearby their neutrinosphere. The state-of-the-art
on the SN dynamics suggests that a significant part of these $\nu_e$ can convert into 
right-handed neutrinos in virtue of the interaction of the electrons and the protons  
flowing with the SN outgoing plasma, whenever the Dirac neutrino magnetic moment be of 
strength $\mu_\nu < 10^{-11} \, \mu_{\rm B}$, with $\mu_{\rm B}$ being the Bohr magneton. 
In the supernova envelope, part of these neutrinos can flip back to the left-handed 
flavors due to the interaction of the neutrino magnetic moment with the magnetic field 
in the SN expanding plasma (Kuznetsov \& Mikheev 2007; Kuznetsov, Mikheev \& Okrugin 
2008), a region where the field strength is currently accepted to be $B \gtrsim 10^{13}$
~G. This type of $\nu$ oscillations were shown to generate powerful gravitational wave 
(GW) bursts (Mosquera Cuesta 2000, Mosquera Cuesta 2002, Mosquera Cuesta \& Fiuza 2004, 
Loveridge 2004). If such double spin-flip mechanism does run into action inside the SN 
core, then the release of both the oscillation-produced $\nu_\mu$s, $\nu_\tau$s and the 
GW pulse generated by the coherent $\nu$ spin-flips provides a unique emission offset 
$\Delta T^{emission}_{\rm GW} \leftrightarrow \nu = 0$ for measuring the $\nu$ travel 
time to Earth. As massive $\nu$s get noticeably delayed on its journey to Earth with 
respect to the Einstein GW they generated during the reconversion transient, then the 
accurate measurement of this time-of-flight delay by SNEWS + LIGO, VIRGO, BBO, DECIGO, 
etc., might readily assess the absolute $\nu$ mass spectrum.

\end{abstract}

%-----------------------------------------------------------------
\section{Introduction}
%-----------------------------------------------------------------

The determination of the absolute values of neutrino masses is
certainly one of the most difficult problems from the experimental
point of view \citep{Bilenky03}. One of the main difficulties of
the issue of determining the $\nu$ masses from solar or
atmospheric $\nu$ experiments concerns the ability of $\nu$
detectors to be sensitive to the species mass-square difference
instead of so doing to the $\nu$ mass itself. In this paper we
introduce a model-independent novel nonpareil method to achieve
this goal. We argue that a highly accurated and largely improved
assessment of the $\nu$ mass-scale can be directly achieved by
measurements of the delay in {\it time-of-flight} between the
$\nu$s themselves and the GW burst generated by the asymmetric
flux of neutrinos undergoing coherent (Pantaleone 1992) helicity 
(spin-flip) transitions
during either the neutronization phase, or the relaxation
(diffusion) phase in the core of a type II SN explosion. Because
special relativistic effects do preclude massive particles of
traveling at the speed of light, while massless do not (the {
{\it graviton}} in this case), the measurement of this $\nu$ time lag
leads to a direct accounting of its mass. We posit from the start
that two bursts of GW can be generated during the PNS {\it
neutronization phase} through spin-flip oscillations: a) one signal 
from the
early conversion of active $\nu$s into right-handed partners, at
density $\rho \sim $ few $10^{12}$ g~cm$^{-3}$, via the interaction  
of the Dirac neutrino magnetic moment (of strength $\mu_\nu < (0.7 -
1.5) \times 10^{-12} \, \mu_{\rm B}$, with $\mu_{\rm B}$ being the 
Bohr magneton) with the electrons and the protons in the SN outflowing 
plasma. Specifically, the neutrino chirality flip is caused by the 
scattering via the intermediate photon (plasmon) off the plasma 
electromagnetic current presented by electrons: $\nu_L e^- 
\longrightarrow \nu_R e^-$, protons: $\nu_L p^+ \longrightarrow \nu_R 
p^+$, etc. b) a second signal in virtue of the reconversion process of 
these sterile $\nu$s back into actives some 
time later, at lower density, via the interaction of the neutrino 
magnetic moment with the magnetic field in the SN envelope. The GW 
characteristic amplitude, which depends directly on the luminosity 
and the mass square-difference of the $\nu$ species partaking in the 
coherent transition \citep{pantaleone92}, and the GW frequency of each 
of the bursts are computed. 
%%%%%%%%%%%%%%%%%%%%%%%%%%  RECONSIDERAR ESTA ARGUMENTACAO !!!!  %%%%%%%%%%%%%%%%%%%%%%%%%%%%%%%%%%%%
%%%% It is shown that they are detectable even if coming from distances halfway to the
%%%% Virgo cluster of galaxies. (Arguments supporting the powerfulness of GW bursts from flavor 
%%%% conversions are provided in Refs.\citep{herman00,herman02,herman03,loveridge}. 
Finally, the time-of-flight delay $\nu \leftrightarrow {\textrm GW}$ that can be
measured upon the arrival of both signals to Earth observatories is then estimated, and the 
prospective of obtaining the $\nu$ mass spectrum from such measurements is discussed.

%%%%%%%%%%%%%%%%%%%%%%%%%%%%%%%%%%%%%%%%%%%%%%%%%%%%%%%%%%%%%%%%%%%%%%%%%%%%%%%%%%%%

\section{Double resonant conversion of neutrinos in supernovae}

\subsection{Interaction of $\nu_L$ Dirac magnetic moment with SN virtual plasmon}

The neutrino chirality conversion process $\nu_L \leftrightarrow \nu_R$ in a supernova 
has been investigated in many papers, see for instance \citep{voloshin88,peltoniemi92,Akhmedov93,amol2000}. 
Next we follow the reanalysis of the double $\nu$ spin flip in supernovae recently 
revisited by Kuznetsov \& Mikheev (2007) and Kuznetsov, Mikheev \& Okrugin (2008), 
who obtained a more stringent 
limit on the neutrino magnetic moment, $\mu_\nu$, after demanding compatibility with the 
SN1987A $\nu$ luminosity. The process becomes 
feasible in virtue of the interaction of the Dirac $\nu$ magnetic moment with a virtual 
plasmon, which can be produced: $\nu_L \longrightarrow \nu_R + \gamma^\star$, and absorbed: 
$\nu_L + \gamma^\star \longrightarrow \nu_R $ inside a SN. Our main goal here is to estimate 
the $\nu_R$ 
luminosity after the first resonant conversion inside the supernova. This quantity is one of 
the important parameters that count to estimate the GW amplitude of the signal generated at 
the transition (see Section \ref{GW-amplitude} below). The calculation of the spin flip rate 
of creation of the $\nu_R$ in the SN core is given by (Kuznetsov \& Mikheev 2007)

\be
L_{\nu_R} \equiv \frac{dE_{\nu_R}}{dt} = V \int^\infty_0 \frac{dn_{\nu_R}}{dE^\prime} E^\prime 
dE^\prime = \frac{V}{2 \pi^2} \int^\infty_0 E^{\prime 3} \Gamma(E^\prime) dE^\prime  \; ,
\label{nuR-luminosity}
\ee

where $\frac{dn_{\nu_R}}{dE^\prime}$ defines the number of right-handed $\nu$s emitted 
in the 1~MeV energy band of the $\nu$ energy spectrum, and per unit time, $\Gamma(E^\prime)$ 
defines the spectral density of the right-handed $\nu$ luminosity, and $V$ is the plasma 
volume. Thus by using the SN core 
conditions that are currently admitted (see for instance \citep{janka-etal2007}): plasma 
volume $V \simeq 4 \times 10^{18}$ cm$^3$, temperature range T = 30-60 MeV, electron 
chemical potential range $\widetilde{\mu_e} = 280-307$ MeV, neutrino chemical potential 
$\widetilde{\mu_\nu} = 160$ MeV\footnote{These conditions could exist in the time interval 
before the first second after the core bounce.}, one obtains

\be
L_{\nu_R} \simeq \left(\frac{\mu_\nu}{\mu_B}\right)^2 (0.4 -2) \times 10^{77} \; ,
{\rm erg \; s}^{-1} 
\label{nuR-effective} 
\ee

which for a $\mu_\nu  = 3 \times 10^{-12}\; \mu_B$ compatible with SN1987A neutrino 
observations, and preserving causality with respect to the left-handed difussion $\nu$ 
luminosity $L_{\nu_R} < L_{\nu_L} \lesssim 10^{53}$~erg s$^{-1}$, renders $L_{\nu_R} = 
4 \times 10^{53}$~erg s$^{-1}$. This constraint is on the order of the luminosities 
estimated in earlier papers  
\citep{herman00,herman02,herman03} to compute the GW amplitude from $\nu$ flavor 
conversions, which were different from the one estimated by \citep{loveridge}. More
remarkable, this analysis means that only $\sim(1-2)\%$ of the total number of $\nu_L$s 
may resonantly convert into $\nu_R$s.

\subsection{Conversion $\nu_R \longrightarrow \nu_L$ in the SN magnetic field }

Kuznetsov, Mikheev \&  Okrugin (2008) have shown that by taking into account the 
additional energy $C_L$, which the left-handed electron type neutrino $\nu_e$ 
acquires in the medium, the equation of the helicity evolution can be written 
in the form \cite{Voloshin1986a,Voloshin1986b,okun1986,Voloshin1986c,okun1988} 

\begin{equation}
{\mathrm i}\,\frac{\partial}{\partial t}
\left(\begin{array}{c} 
\nu_R \\ \nu_L 
\end{array} 
\right)
=
\left[\hat E_0 +
\left( 
\begin{array}{cc} 
0 & \mu_\nu B_{\perp} \\ \mu_\nu B_{\perp} & C_L
\end{array} 
\right) 
\right]
\left( 
\begin{array}{c} 
\nu_R \\ \nu_L 
\end{array} 
\right) \,,
\label{eq:evolution} 
%  \end{equation} %where
%  \begin{equation}
\hskip 0.3 truecm \because \hskip 0.3 truecm C_L = \frac{3 \, G_{\mathrm F}}{\sqrt{2}} \, 
\frac{\rho}{m_N} \left( Y_e + \frac{4}{3} \, Y_{\nu_e} - \frac{1}{3} \right) \,.
%%%   \label{eq:C_L}
\end{equation}

Here, the ratio $\rho/m_N = n_B$ is the nucleon density, while $Y_e = n_e/n_B = n_p/n_B, \, 
Y_{\nu_e} = n_{\nu_e}/n_B$, $n_{e,p,\nu_e}$ are the densities of electrons, protons and 
neutrinos, respectively. $B_{\perp}$ is the transverse component of the magnetic field 
with respect to the $\nu$ propagation direction, and the term $\hat E_0$ is proportional 
to the unit matrix, however, it is not crucial for the analysis below. 

%%%%%%%%%%%%%%%%%%%%%%%%%%%%%%%%%%%%%%%%%%%%%%%%%%%%%%%%%%%%%%%%%%%%%%%%%%%%%%%%%%%%%%%%%%%%%%%%%%%

As pointed out by Kuznetsov, Mikheev \& Okrugin (2008), the additional energy $C_L$ of 
left-handed $\nu$s deserves a special analysis. It is remarkable that the possibility 
exists for this value to be zero just in the region of the supernova envelope (SNE) we 
are interested in. And in turn this is the condition of the resonant transition $\nu_R 
\to \nu_L$. As the $\nu$ density in the SNE is low enough, one can neglect the value 
$Y_{\nu_e}$ in the term $C_L$, which gives the condition for the resonance in the form 
$Y_e = 1/3$. (Typical values of $Y_e$ in SNE are $Y_e \sim 0.4-0.5$, which are rather 
similar to
those of the collapsing matter). However, the shock wave causes the nuclei dissociation and 
makes the SNE material more transparent to $\nu$s. This leads to the proliferation of matter 
deleptonization in this region, and consequently to the so-called ``short'' $\nu$ outburst. 
According to the SN state-of-the-art, a typical gap appears along the radial distribution 
of the parameter $Y_e$ where it can achieve values as low as $Y_e \sim 0.1$ (see 
\cite{mezzacappa}, and also Fig. 2 
in Kuznetsov, Mikheev \& Okrugin 2008, and references therein). Thus, a transition region
unavoidably exists where $Y_e$ takes the value of $1/3$. It is remarkable that only one 
such point appears where the $Y_e$ radial gradient is positive, i.e, $\mathrm{d}Y_e/ 
\mathrm{d}r > 0$. Nonetheless, the condition $Y_e = 1/3$ is the necessary but yet not 
the sufficient one for the resonant conversion $\nu_R \to \nu_L$ to occur. It is also required 
to satisfy the so-called adiabatic condition. This means that the diagonal element $C_L$ 
in the equation (\ref{eq:evolution}), at least, should not exceed the nondiagonal element 
$\mu_\nu B_{\perp}$, when the shift is made from the resonance point at the distance 
of the order of the oscillation length. This leads to the condition \cite{voloshin88} 

\begin{eqnarray}
\mu_\nu B_{\perp} \gtrsim \left( \frac{\mathrm{d} C_L}{\mathrm{d} r} \right)^{1/2} 
\simeq \left( \frac{3 \, G_{\mathrm F}}{\sqrt{2}} \, \frac{\rho}{m_N} \, 
\frac{\mathrm{d} Y_e}{\mathrm{d} r}\right)^{1/2}\; .
\label{eq:res_cond}
\end{eqnarray}

And values of these typical parameters inside the considered region are: 
$\frac{\mathrm{d} Y_e}{\mathrm{d r}} \sim 10^{-8} \, \mbox{cm}^{-1} \,, 
\quad \rho \sim 10^{10} \, \mbox{g} ~\mbox{cm}^{-3}$. Therefore, the magnetic 
field strength that realizes the resonance condition reads

\begin{eqnarray}
B_{\perp} \gtrsim 2.6 \times 10^{14} \mbox{G} \left( \frac{10^{-12} \mu_{\rm B}}{\mu_\nu} 
\right) \left(\frac{\rho}{10^{10} \mbox{g} ~ \mbox{cm}^{-3}}\right)^{1/2}
\left( \frac{\mathrm{d} Y_e}{\mathrm{d} r} \times 10^8 \, \mbox{cm} \right)^{1/2} .
\label{eq:res_cond_B}
\end{eqnarray}

Thus, one can conclude that the analysis performed above shows that the Dar' scenario of the 
double conversion of the neutrino helicity  \citep{dar1987}, $\nu_L \to \nu_R \to 
\nu_L$, can be realized whenever the neutrino magnetic moment is in the interval $
10^{-13} \, \mu_{\rm B} < \mu_\nu < 10^{-12} \, \mu_{\rm B}$, and when the strength
of the magnetic field reaches $\gtrsim 10^{14}$~G \citep{kusenko04} in a region $R$ 
between the neutrinosphere 
$R_\nu$ and the shock wave stagnation radius $R_s$, where $R_\nu < R < R_s$.\footnote{
These kind of magnetic field strengths have been extensively said to be reached after the SN 
core-collapse to form just-born pulsars (magnetars), in the central engines of gamma-ray burst 
outflows, and during the quantum-magnetic collapse of new-born neutron stars, etc.} Thus, the
$\nu_L$ luminosity during this stagnation time, $\Delta T_s \simeq 0.2-0.4$ sec, is
$L_{\nu_L} \simeq 3 \times 10^{53}$ erg~s$^{-1}$, as the conservation law allows to 
expect for $\mu_\nu < 10^{-12} \, \mu_{\rm B}$. Once having all these parameters in hand 
one can then proceed to compute the corresponding GW signal from each of the $\nu$ resonant 
spin-flip transitions.

%%%%%%%%%%%%%%%%%%%%%%%%%%%%%%%%%%%%%%%%%%%%%%%%%%%%%%%%%%%%%%%%%%%%%%%%%%%%%%%%%%%%%%%%%%%%%%%%%%%

%-----------------------------------------------------------------
\section{\label{GW-amplitude} $\nu$ oscillation-driven GW during SN neutronization}
%-----------------------------------------------------------------

The characteristic GW amplitude of the signal produced by the $\nu$s outflow can be 
estimated by using the general relativistic quadrupole formula \citep{burrows96}

 \begin{equation}
 h{^{TT}_{ij}}(t)  = \frac{4 G}{c^4 D}  \int^t_{-\infty}
 \alpha(t') L_\nu(t') ~ dt' \; \; e_i \otimes e_j \; , \longrightarrow
 h \simeq \frac{4 G}{c^4 D} \; \alpha \; \Delta L_{\nu} 
\; \Delta T_{\rm \nu_{f_L} \to \nu_{f_R}} 
 \label{GW-AMPLITUDE}
 \end{equation}

where $D$ is the source distance, $L_\nu(t)$ the total $\nu$
luminosity, $e_i \otimes e_j$ the GW polarization tensor, the
scripts $TT$ stand for transverse-traceless part, and finally,
$\alpha(t)$ is the instantaneous quadrupole anisotropy. Above 
we estimated the $\nu_R$ luminosity, next we estimate the degree
of asymmetry of the proto-neutron star through the anisotropic 
parameter $\alpha$, and the timescale $\Delta T_{\rm \nu_{f_L} 
\to \nu_{f_R}}$ for the resonant transition to take place, as discussed above.

To estimate the star asymmetry, let us recall that the resonance 
condition for the transition $\nu_{eL}\to \nu_{\mu R}$ is given 
by (at the resonance ${\bar r}$)

\begin{equation} \label{res}
V_{\nu_e}({\bar r})+{\bsymbol{B}}({\bar r})\cdot {\hat {\textbf p}}
- 2\delta c_2=0\,.
\end{equation}

\begin{figure}
\centering
%%%  \framebox{    {!}
\resizebox{7.4cm}{!}{
\includegraphics{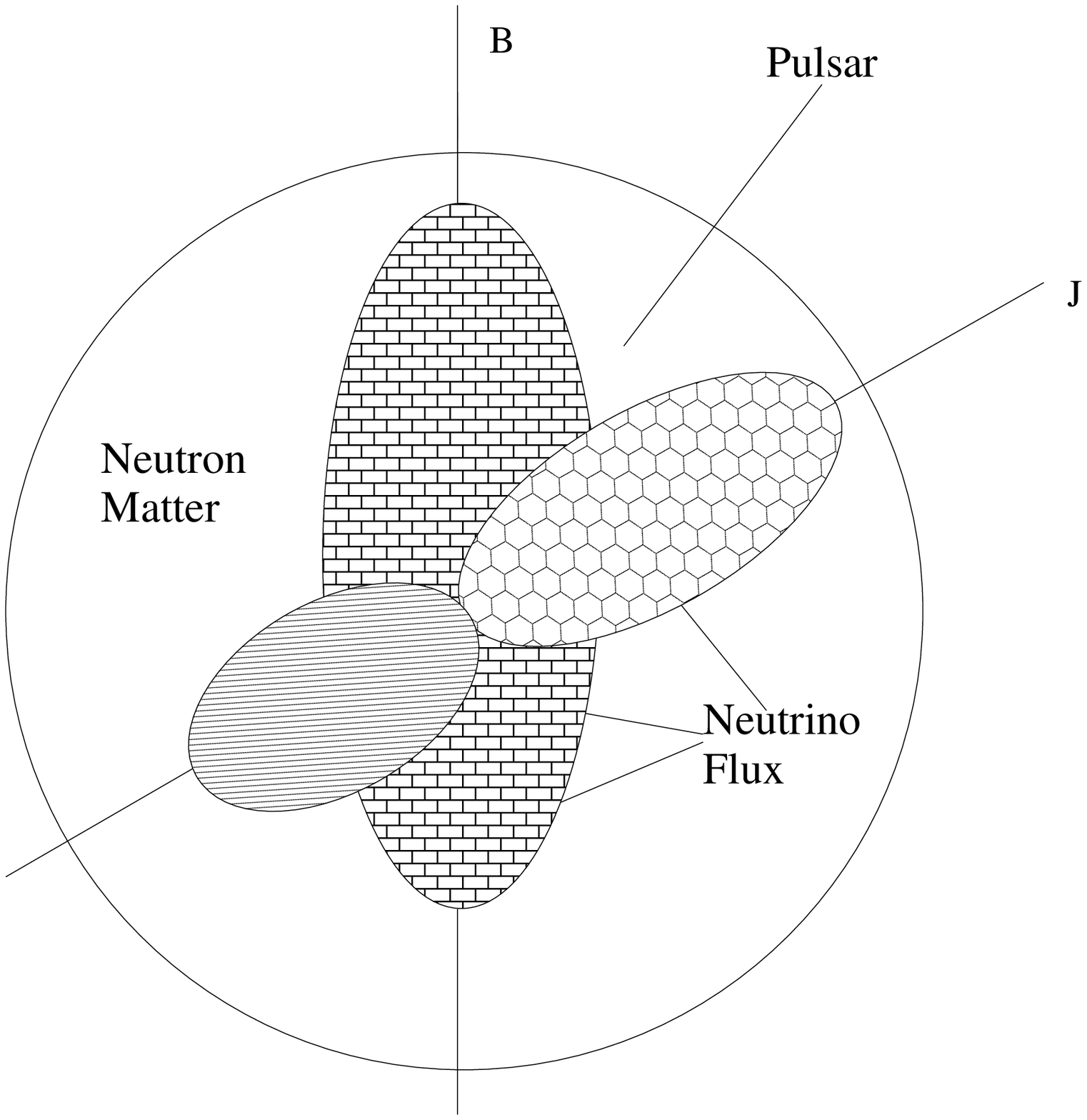} }
\caption{Illustration of the combined effect of the $\nu$ spin 
coupling to the star magnetic field and rotation. [Taken 
from H. J. Mosquera Cuesta \& K. Fiuza, Eur. Phys. Journ. C 35, 543 (2004)]. }
\label{anisotropy}
\end{figure}

Thus, the proto-neutron star (PNS) magnetic field vector 
${\bsymbol{B}}$ in (\ref{res}) distorts the surface of resonance
due to the relative orientation of ${\textbf p}$ with respect to 
${\bsymbol{B}}$ (see vector $\vec{B}$ in Fig. \ref{anisotropy}). 
The {\it deformed} surface of resonance can be
parameterized as $r(\beta)={\bar r}+\varrho\cos\beta$, where
$\varrho (<{\bar r})$ is the radial deformation and $\cos \beta =
\hat{{\bsymbol{B}}} \cdot {\hat {\textbf p}}$. The deformation 
enforces a non-symmetric outgoing neutrino flux, i.e., the net flux of
neutrinos emitted from the upper hemisphere is different from the
one emitted from the lower hemisphere (see Fig. \ref{anisotropy}). 
Therefore, a geometrical definition of the quadrupole anisotropy 
can be: $\alpha = \frac{S_+ - S_-}{S_+ + S_-}$, where $S_\pm$ is 
the area of the up/down hemisphere, whence one
obtains $\alpha \simeq\varrho/{\bar r}$ \footnote{A detailed
analysis of the asymmetry parameter $\alpha$ requires to study its
time evolution during the SN collapse. Such a task goes beyond the
aim of this paper. Working in stationary regime, we may assume
$\alpha$ constant (see \citep{burrows96,burrows95,zwerger,vanputten02}).}.
The anisotropy of the outgoing neutrinos is also related to the
energy flux ${\textbf F}_s$ emitted by the PNS, and in turn to the
fractional momentum asymmetry $\Delta |\vec{p}|/|\vec{p}|$
\citep{kusenko,zanella,lambiaseMNRAS,herman03}. To compute 
${\textbf F}_s$, one has to take into account the structure of the flux at
the resonant surface, which acts as an effective emission surface,
and the $\nu$ distribution in the diffusive approximation
\citep{zanella}. As a result, one gets $\frac{\Delta |\vec{p}|}{
|\vec{p}|} = \frac{1}{6} \frac{\int_0^\pi {\textbf F}_s\cdot {\textbf u}\,
dS}{\int_0^\pi {\textbf F}_s \cdot {\textbf n}\, dS} \simeq
\frac{2\varrho} {9{\bar r}}$ (${\textbf n}$ is a unit vector normal to
the resonance surface, and ${\textbf u} = {\hat{\bsymbol{B}}}/|{\hat
{\bsymbol{B}}}|$)\footnote{To compute $\Delta |\vec{p}|/|\vec{p}|$ 
one uses the standard resonance condition $V_\nu = 2\delta c_2$ (see
\citep{zanella} for details). According to \citep{mezzacappa},
during the first (10-200) ms, $Y_e$ may assume values $\simeq
1/3$ so that $V_{\nu_e}\sim (3Y_e-1)$ is suppressed by several
order of mangitude. At $\sim 10$ ms, $\rho\sim
10^{12}$~gr~cm$^{-3}$, $r\sim 50$~km, and for $|\vec{p}| \sim$ 10 MeV, the
resonance condition leads to a range for $\Delta m^2\cos 2\theta$
consistent with solar (or atmospheric) neutrinos data.}. An
anisotropy of $\sim 1\%$ would suffice to account for the observed
pulsar kicks \citep{kusenko,loveridge,herman00,herman02}, hence
$\alpha\simeq 0.045\sim {\cal O} (0.01)-{\cal O}(0.1)$, which is
consistent with numerical results of \citep{burrows96,Janka}. 
Finally, the conversion probability is $P_{\nu_{eL}\to \nu_{\mu R}} = 1/2-1/2\cos
2{\tilde \theta}_i \cos 2 {\tilde \theta}_f$ \citep{okun1986,okun1988}, where 
${\tilde \theta}$ is defined as

\be
\tan 2 {\tilde \theta}(r)=2\mu_\nu B_\perp/({\bsymbol{B}}\cdot {\hat 
{\textbf p}} + V_{\nu_e} - 2\delta c_2)\,.
\ee

${\tilde \theta}_i={\tilde \theta}(r_i)$ and ${\tilde
\theta}_f={\tilde \theta}(r_f)$ are the values of the mixing angle
at the initial point $r_i$ and the final point $r_f$ of the
neutrino path\footnote{By using the typical values $B\gtrsim
10^{10}$~G, $\mu_{\nu}\lesssim 9\,\, 10^{-11}\mu_B$, and the
profile $\rho\simeq \rho_{core} (r_c/r)^3$ for $r\gtrsim r_c$
($r_c\sim 10$~km is the core radius and $\rho_{core}\sim
10^{14}$~gr~cm$^{-3}$), one can easily verify that the adiabatic
parameter $\gamma\equiv \frac{2(\mu_{\nu} B_\perp)^2}{\delta \pi
|\rho'/\rho|}> 1$ at the resonance ${\bar r}$.}.

%%%%%%%%%%%%%%%%%%%%%%%%%%%%%%%%%%%%%%%%%%%%%%%%%%%%%%%%%%%%%%%%%%%%%%%%%

Meanwhile, the average timescale of this first $\nu$ spin-flip conversion 
is \citep{dar1987,voloshin88} 

\begin{equation}
\Delta T_{\rm \nu_{f_L} \to \nu_{f_R}} = \left(\frac{\mu_B}{\mu_\nu} 
\right)^2  \left\{ \frac{m_e^2} {\pi \alpha^2_{\rm fsc} (1 + \langle 
Z \rangle) Y_e} \right\} \left[\frac{m_p}{\rho}\right]\,,
\label{timescale}
\end{equation}

where $\langle Z \rangle\sim {\cal O}(1-30)$ is the average
electric charge of the nuclei, and $\alpha_{\rm fsc}$ the fine structure
constant. Using the current bounds on the neutrino magnetic
moment $\mu_\nu\lesssim 3\times 10^{-12}\mu_B$, $Y_e\simeq 1/3$,
$\langle Z \rangle\sim 10$, $\rho\sim 2\times 10^{12}$gr/cm$^3$, and
$\alpha \sim 0.04$, it follows $\Delta T_{\rm \nu_{fL} \to 
\nu_{fR}} \simeq (1-10) \times 10^{-2}$~sec (parameters have been chosen
from SN simulations evolving the PNS in time scales of $\sim 3$~ms
around core-bounce
\citep{mayle87,walker1987A,burrows96,mezzacappa,vanputten02,arnaud02,beacom01}).
In such a case, the above timescale suggests that the GW burst would be as long 
as the expected duration of the pure neutronization phase itself, i.e., $\Delta
T_{\rm Neut} \sim (10-100)$~ms, according to most SN analysis and models
\citep{mayle87,walker1987A,burrows96,mezzacappa,vanputten02,arnaud02,beacom01},
with the maximum GW emission taking place around $\Delta T^{\rm
max}_{\rm Neut} \sim 3$~ms
\citep{vanputten02,arnaud02,herman00,herman02,herman03}. Hence,
the out-coming GW signal will be the evolute (linear
superposition) of all the coherent $\nu_{eL} \to \nu_{\mu,\tau R}$
oscillations taking place over the neutronization transient, in
analogy with the GW signal from the collective motion of neutron
matter in a just-born pulsar. This implies a GW frequency: $f_{\rm
GW} \sim \Delta T_{\rm Neut}\sim 100$~Hz, for the overall GW emission,
and $f_{\rm GW} \sim 1/\Delta T^{\rm max}_{\rm Neut} \sim 330$~Hz at
its peak. Meanwhile, according to our probability discussion above
about (1-2)\% of the total $\nu$s released during the
SN neutronization phase may oscillate
\citep{voloshin88,peltoniemi92,Akhmedov93,amol2000} carrying away an
effective power: $L_\nu = 3\times 10^{54-53}$ erg s$^{-1}$, i.e.,
$0.01 \times 3\times 10^{53}$~erg, emitted during $\Delta T_{\rm Neut}
\sim (10-100)$~ms (this is similar to the upper limit computed in
Ref.\citep{peltoniemi92}: $L_\nu =(2\div 10) \times 10^{53}
\left(\frac{\mu_{\nu_e}}{10^{-12} \mu_B}\right)$~erg~s$^{-1}$).
Moreover, as is evident from Eq.(\ref{GW-AMPLITUDE}), the GW
amplitude is a function of the helicity-changing $\nu$ luminosity,
i.e., $h = h(L^{\nu_{eL} \to \nu_{\mu,\tau R} }_{\rm max})$. The
$\nu$ luminosity itself depends on the probability of conversion
\citep{peltoniemi92,herman00,herman02,herman03,loveridge}, i.e., 
$L^{\nu_{eL} \to \nu_{\mu,\tau R} }_{\rm max} = (P_{\nu_{eL} \to
\nu_{\mu,\tau R}}) L^\nu_{\rm total}$.

\begin{figure}
\centering
%%%  \framebox{    {!}
\resizebox{7.4cm}{!}{
\includegraphics{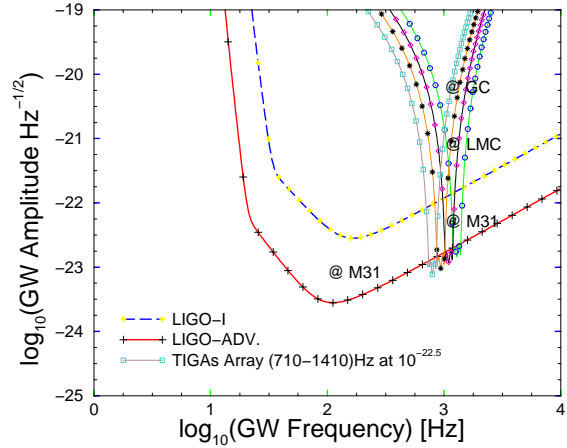} }
%%%%  }
%%%% \vspace{302pt}
\caption{Characteristics ($h^{({\nu_{fL} \to \nu_{f'R}})}$,
$f_{\rm GW}$) of the GW burst generated via the $\nu$ spin-flip
oscillation mechanism vs. detectors {\it noise} spectral density.
For sources at either the GC or LMC the pulses will be detectable
by LIGO-I, VIRGO. To distances $\sim$ 10~Mpc (farther out the
Andromeda galaxy) such a radiation would be detectable by Advanced
LIGO, VIRGO. Resonant gravitational-wave antennas, tuned at the
frequency interval indicated, could also detect such events.
Highlighted is the GW signal of a SN neutronization phase at
Andromeda, which would have a frequency: $f_{\rm GW} \sim
100$~Hz.}
\label{GW-sensitivity}
\end{figure}

The characteristic GW strain (per $\sqrt{\textrm Hz}$) from the
outgoing flux of spin-flipping (first transition) $\nu$s is

\begin{equation}
\label{herman03A}
h^{({\nu_{fL} \to \nu_{f'R} })} \equiv h \simeq 1.1 \times 10^{-23}
~[{\textrm Hz}^{-1/2}] \; \frac{P_{\nu_{fL} \rightarrow \nu_{f' R}}
}{0.01} \; \frac{ L^{\rm total}_\nu}{ 3\times 10^{54}~\frac{\rm erg}
{\rm s}} \; \frac{2.2 ~\rm Mpc}{D} \; \frac{\Delta T }{10^{-1}~ \rm
s} \; \frac{\alpha} {0.1}\,,
\end{equation}

for a SN exploding at a fiducial distance of 2.2~Mpc, e.g., at
the Andromeda galaxy (see Table~I\footnote{The mass eingenstates 
listed are masses supposed to be estimated throughout 
the $\nu$ detection in a future SN event, not the mass constraints already 
stablished from solar and atmospheric neutrinos, the expected time-delay of 
which is straightway computable. If a non-standard mass eigenstate is detected,
then one can use the {\it see-saw} mechanism to infer the remaining part of 
the spectrum.}). The GW strain in this
mechanism (see Fig.~\ref{GW-sensitivity}) is several orders 
of magnitude larger than in the SN $\nu$ diffusive escape
\citep{burrows96,Janka,arnaud02,loveridge} because of the huge
$\nu$ luminosity the $\nu$ oscillations provide by cause of being
a highly coherent process
\citep{pantaleone92,herman00,herman02,herman03}. This makes it
detectable from very far distances. These GW signals are right in the
band width of highest sensitivity [10-300] Hz of most ground-based
interferometers.

%%%%%%%%%%%%%%%%%%%%%%%%%%%%%%%%%%%%%%%%%%%%%%%%%%%%%%%%%%%%%%%%%%%%%%%%%%%%

Spin flavor oscillations $\nu_{eL} \to \nu_{\mu R}$, which according to 
the state-of-the-art of SN dynamics do take place during the {\it 
neutronization phase} of core collapse supernovae
\citep{mayle87,walker1987A,voloshin88,amol2000,Kuznetsov2007}, 
allow from one side to release powerful GW bursts (according to
(\ref{GW-AMPLITUDE})), and from the other side to generate, over a
timescale given by (\ref{timescale}), a stream of $\nu_{\mu R}$s.
The latter would in principle escape from the PNS were not by the
appearance of several resonances that catch them up before
\citep{voloshin88,peltoniemi92,Akhmedov93}. There were no such a
resonance the $\nu_{fL} \to \nu_{f'R}$ scenario would leak away
all the binding energy of the star leaving no energy at all for
the left-handed $\nu_L$s that are said to drive the actual SN 
explosion; and to allow us to observe them during SN1987A.
A new resonance may occur at $\bar{r} \gtrsim 100$~km from the
center which converts back $\sim 90-99\%$ of the
spin-flip-produced $\nu_R$s into $\nu_L$ ones
\citep{voloshin88,akhmedov88,peltoniemi92,Akhmedov93,athar95}. As
discussed in these papers, in fact, in the outer layer of the
supernova core the amplitude of the coherent weak interaction of
$\nu_L$ with the PNS matter ($V_{\nu_e}$) can cross smoothly
enough to ensure adiabatic resonant conversion  of $\nu_{fR}$ into
$\nu_{fL}$\footnote{The cross level condition once again involves
the terms ${\bsymbol{B}}\cdot {\hat {\textbf p}}$. Nevertheless, 
at that point the deformation of the resonance surface may be
neglected, whence no relevant GW burst is expected (yet $\rho$ is
quite low).}. Following \citep{mezzacappa} the region where
$V_{\nu_e}=0$ as $Y_e=1/3$ corresponds to a post-bounce timescale
$\sim 100$~ms and radius $\sim 150$~km at which the $\nu$
luminosity is $L_\nu \sim 3 \times 10^{52}$ erg/s, and the matter
density $\rho \sim 10^{10}$g/cm$^{3}$. There the adiabaticity
condition demands $B_\perp \gtrsim 10^{10}$~G for the $\mu_\nu$
quoted above (such a field is characteristic of young pulsars).
This reverse transition (rt) should resonantly produce an
important set of ordinary (muon and tau) $\nu_L$s, which would
find far from their own $\nu$-sphere and hence can stream-away
from the PNS. Whence a second GW burst with characteristics: $h
\simeq 1\times 10^{-23}$ ~Hz$^{-\frac{1} {2}}$ for $D = 2.2$~Mpc,
and $\Delta T_{rt} \simeq 1.4~$s is  released in
this region. Notice that this $h$ is similar to the one for the
first transition despite the $\nu$ luminosity is lower. A feature 
that make it similar to the GW memory property of the $\nu$-driven 
signal, i.e., time-dependent strain amplitude with average value nearly 
constant \citep{burrows96}. To obtain this result Eqs.(\ref{timescale},
\ref{herman03A}) were used. Wherefore, the GW frequency $f_{\rm
GW} \sim 1/\Delta T_{rt} \sim 0.7$~Hz falls in the low frequency
band and could be detected by the planned BBO and DECIGO GW
interferometric observatories. Notice also that the time lag for
the event at LIGO, VIRGO, etc., and the one at BBO, DECIGO is then
about 100~ms. It is this transition what defines the offset to
measure the time-of-flight delay since both $\nu_{\mu,\tau}$ and
GW free-stream away from the PNS at this point.

%-----------------------------------------------------------------
\section{Time-of-flight delay $\nu \leftrightarrow {\textrm GW}$}
%-----------------------------------------------------------------

The $\nu \leftrightarrow$ GW time delay from $\nu$ oscillations in
SN promises to be an inedit procedure to obtain the $\nu$ mass
spectrum. Provided that Einstein's gravitational waves do propagate at
the speed of light, the GW burst produced by spin-flip
oscillations during the neutronization phase will arrive to GW
observatories earlier than its source (the massive $\nu$s from the
second conversion) will get to $\nu$ telescopes.

%%%%%%%%%%%%%%%%%%%%%%%%%%%%%%%%%%%%%%%%%%%%%%%%%%%%%%%%%%%%%%%%%%%

As pointed out earlier, the mechanism to generate GWs at the
instant in which the second transition $\nu_{f'R} \to \nu_{fL}$
takes place can by itself define a unique emission offset, $\Delta
T^{emission}_{\textrm GW \leftrightarrow \nu} = 0$, which makes it
possible a cleaner and highly accurate determination of the $\nu$
mass spectrum by ``following'' the GW and neutrino propagation to
Earth observatories. The time lag in arrival is \citep{beacom01} 
\vskip -0.4 truecm

\begin{equation}
 \Delta T^{arrival}_{\rm GW \leftrightarrow \nu} \simeq 0.12 \,
{\rm s} \left[\frac{D}{2.2\; \rm Mpc}\right] \left[\frac{m_{\nu}}
{0.2 \;\rm eV}\frac{10\, \rm MeV}{|\vec{p}|}\right]^2\,. 
\label{delay}
 \end{equation}

%-----------------------------------------------------------------
\section{Discussion}
%-----------------------------------------------------------------

In most SN models
\citep{burrows96,mezzacappa,beacom01,vanputten02} the
neutronization burst is a well characterized process of {\it
intrinsic} duration $\Delta T \simeq 10$~ms, with its maximum
occurring within $3.5 \pm 0.5$~ms after core collapse
\citep{mayle87,walker1987A,vanputten02,burrows96}. This timescale
relates to the detectors approximate sensitivity to $\nu$ masses
beyond the mass limit
 \begin{equation}
 m_\nu > 6.7 \times 10^{-2}~{\rm eV} \left[{\frac{2.2 ~\rm
Mpc}{D}} \,{\frac{\Delta T}{\rm 10~ms}}\right]^\frac{1}{2}
\left(\frac{|\vec{p}|}{10~\rm MeV}\right). 
\label{mass-sensitivity}
 \end{equation}
%\floatfix
A threshold in agreement with the current bounds on $\nu$ masses
\citep{fukuda98}.

\begin{table*}
%\hline
%\hline
\caption{\label{tbl-2}{Time delay between GW and ($|\vec{p}| = 10$~MeV) $\nu$
bursts from a SN neutronization, as a function of $\nu$ mass and distance. } }
\begin{tabular}{ccccccc}
\\
\hline\hline
 \multicolumn{1}{c}{ } & \multicolumn{1}{c}{$\nu$
flavor } & \multicolumn{1}{c}{{$\nu$ Mass } } &
\multicolumn{1}{c}{ GC } & \multicolumn{1}{c}{ LMC } &
\multicolumn{1}{c} { M31 } &
\multicolumn{1}{c} { Source } \\
%%%\hline
\vspace{3pt} 
{   } & {   } & { [eV] } & { [10 kpc] } & { [55 kpc]} & { [2.2 Mpc] } & { [11 Mpc] } \\
\hline \hline
{   }  & { $\nu_1$ }  & { $10^{-3}$  } & { $5.15\times 10^{-9} $ }
& { $2.83\times 10^{-8} $ } & { $1.13 \times 10^{-6}$ } & $5.66\times 10^{-6}$
\\ 
{ {$\Delta T^{arrival}_{\rm GW \leftrightarrow \nu }$ } } & { $\nu_2$ }
& { $1.0$  } &  { $5.15\times 10^{-3} $ } & { $2.83\times 10^{-2}$ } & { 1.13 } &
5.66 \\
{ [s]  } & { $\nu_3$ }  & { $2.5$ } &  { $0.32 $  } & { $1.7$ } & { 68.8 } & 344.0  \\
\hline \hline
\end{tabular}
\end{table*}

Nearby SNe will somehow be seen. Apart from GW and $\nu$s,
$\gamma$-rays, x-rays, visible, infra-red, or radio signals will
be detected. Therefore, their position on the sky and distance
($D$) may be determined quite accurately, including; if far from
the Milky Way, their host galaxy \citep{Ando05}. Besides, the
Universal Time of arrival of the GW burst to three or more
gravitational radiation interferometric observatories or resonant 
detectors will be precisely established \citep{schutz86,arnaud02}. The
uncertainty in the GW timing depends on the signal-to-noise ratio
($SNR$) as $\Delta T ({\textrm GW}|_{ D = 10 \textrm kpc}) \sim 1.45
\tau/SNR \sim 0.15$~ms, with $\tau \sim 1$~ms the {\it rms} width
of the main GW peak \citep{arnaud02}. Meanwhile, the type of $\nu$
and its energy and Universal Time of arrival to $\nu$ telescopes
of the SNEWS network will be highly accurately measured
\citep{Antonioli99,beacom99}. The $\nu$ timing uncertainty is
$\Delta T_\nu^{\rm max} = \sigma_{flash}(N_\nu)^{-1/2}$, with
$\sigma_{flash} \sim (2.3 \pm 0.3)$~ms, and $N_\nu$ the event
statistics (proportional to $D$). This leads to the SN
distance-dependent uncertainty in the $\nu$ mass:  $\delta m^2_\nu
\propto \Delta T_\nu^{\rm max}/D \sim 0.5-0.6$~eV$^2$
\citep{arnaud02}, which implies a $m_\nu \sim 7 \times
10^{-1}$~eV, which is consistent with our previous estimate
(\ref{mass-sensitivity}). Hence, those $\nu$s and their spin-flip
conversion signals must be detected.

Therefore, the left-hand-side of Eq.(\ref{delay}), i.e., the
time-of-flight delay $\Delta T_{\rm GW \leftrightarrow \nu}$, will
be measured with a very high accuracy. With these quantities a
very precise and stringent assessment of the absolute $\nu$
mass-eigenstate spectrum will be readily set out by means not
explored earlier in astroparticle physics: An inedit technique
involving not only particle but also GW astronomy. For instance, 
at a 10 kpc distance, e.g., to the galactic center (GC in Fig. 
\ref{GW-sensitivity}), the
resulting time delay should approximate: $ \Delta T_{\rm GW
\longleftrightarrow \nu} = 5.2 \times 10^{-3} \;$s, for a flavor
of mass $m_{\nu} \leq 1~{\rm eV}$ and $|\vec{p}| \sim 10$~MeV. A SN event
from the GC or Large Magellanic Cloud (LMC) would provide enough
statistics in SNO, SK, etc. $\sim 5000-8000$ events, so as to
allow for the definition of the $\nu$ mass eigenstates
\citep{beacom01}. Farther out $\nu$ events are less promising in
this perspective, but we stress that one $\nu$ event collected by 
the planned Megaton $\nu$ detector, from a large distance source, 
may prove suffice, see further arguments in \citep{Ando05}.

%-----------------------------------------------------------------
\section{Summary}
%-----------------------------------------------------------------

In this paper, it has been emphasized that knowing with enough
accuracy the $\nu$ absolute mass-scale would turn out in a
fundamental test of the physics beyond the standard model of
fundamental interactions. In virtue of the very important 
two-step mechanism of $\nu$ spin-flavor conversions in supernovae,
very recently revisited by Kuznetsov, Mikheev \& Okrugin (2008),
we suggest that by combining the detection of the GW signals 
generated by those oscillations and the $\nu$ signals collected 
by SNEWS from the same SN
event, one might conclusively assess the $\nu$ mass spectrum. In
special, sorting out the neutronization phase signal from both the
$\nu$ lightcurve and the second peak in the GW waveform (with its
memory-like feature \citep{burrows96}) might allow to achieve this 
goal in a nonpareil fashion.

HJMC {\it is fellow of the Funda\c c\~ao de Amparo \`a Pesquisa do 
Estado de Rio de Janeiro, FAPERJ, Brazil}. G.L. {\it aknowledges 
the support provided by MIUR through PRIN Gravitational Lensing, 
Dark Matter and Cosmology}

%-----------------------------------------------------------------
%%% \begin{acknowledgements}
%%% The authors thank S. Mohanty and J. F. Nieves for valuable discussions.
%%% \end{acknowledgements}

%
------------------------------------------------------------------

\bibliographystyle{aa}

\end{document}